\documentclass[aps,prl,preprint,groupedaddress]{revtex4}

\usepackage{graphics}
\usepackage{rotating}

\begin{document}

\title{Calculation of the two--photon decay width of the $f_0(980)$
scalar meson}

\author{R. H. Lemmer$^{\ddag}$}

\affiliation{Physik--Department der Technischen Universit\"at M\"unchen,
D--85747, Garching, Germany}

\date{\today}

\begin{abstract}
   The applicability of the quasi--static approximation
   for calculating  the two--photon annihilation rate of the
   scalar $f_0(980)$ meson envisaged as a
     $K\bar K$ molecule is critically re--examined.
 It is shown that the validity of this approximation depends on
 the detailed interplay between the momentum dependence  of the annihilation
 amplitude and the momentum space transform of the bound state wavefunction
  of the annihilating pair. The approximation becomes invalid
 when these two scales of variation are similar. An improved method of
 calculation based on the inclusion of electromagnetic corrections to the kernel
 of the Bethe--Salpeter equation for the interacting $K\bar K$ pair
 is outlined to cover this case and applied to re-evaluate the two--photon decay
 width for $f_0(980)$ in a one boson exchange model for the
 interkaon interaction.  The corrections  are significant and result in a much better
 agreement with experiment.


\vspace{10cm}
\noindent
($^\ddag$) Permanent address: School of Physics,
 University of the Witwatersrand, Johannesburg,\\ Private Bag 3,
 WITS 2050, South Africa.

\vspace{1cm}
Corresponding author: R H Lemmer

Electronic address: rh$_-$lemmer@mweb.co.za

\end{abstract}
\maketitle

\newpage

 \subsection{1. Introduction}

  The structure of the lowest mass scalar meson $f_0(980)$ has been under debate
    for some time. As possible candidates a $q\bar q$ state \cite{LM00},
  a $q^2\bar q^2$ state \cite{NNA89},
   a $ K\bar K$ molecule \cite{WIsgur82}, or perhaps some combination of these
   structures \cite{KLS04} have been suggested. However,
   the large two--pion decay width
   predicted for
   the first option, $\Gamma(q\bar q\to \pi\pi)\sim 500$ MeV from
   flux tube--breaking \cite{RKNI87} to $\sim 660 $ MeV using current algebra \cite{RT02},
   does not favor a pure $q\bar q$ configuration 
   in view of
   the typical experimental  width \cite{PDG06} of  $\sim (40 - 100)$ MeV  for
   the $f_0(980)\to\pi\pi$ decay.

 On the other hand  Barnes \cite{TB85}, in an earlier paper, advanced cogent arguments in
   support of the  molecular picture, using various 
  experimental data to obtain an estimate of the two--photon
  decay width of the
  $f_0(980)$ that agrees  qualitatively with experiment when this scalar meson is
  viewed as a $K\bar K $ molecule. As in the case of positronium
  \cite{MAS75}, Barnes' calculation relies on the validity
  of the quasi-static, or wavefunction at contact approximation, where the annihilation
  rate is computed in Born approximation from the free annihilation
  cross--section 
    multiplied by a kaon current that is proportional to the probability of having
  a bound $K^+K^-$ pair at the origin. The latter quantity in turn depends on the
  assumed two--body potential that produces the molecular binding.
  In \cite{TB85} this was taken to have a gaussian shape, with a depth and range
  typical of the interkaon Weinstein--Isgur potential \cite{WIsgur82}. 

  Repeating the same calculation for the one boson exchange potential
  developed in \cite{KLS04}
  one finds that the
  two--photon width comes out 
  an order of magnitude larger than for the Barnes' potential, although the two--pion width 
   calculated
  in the same quasi--static approximation agrees with experiment.
      We have  therefore re--examined the validity of the quasi--static
 approximation 
   by implementing
 a more complete
 calculation using the Bethe--Salpeter equation.
 It is shown that the applicability of this approximation depends on
 the detailed interplay between the momentum dependence scales of the annihilation
 amplitude and the momentum space transform of the bound state wave function
 $\psi(r)$ of the annihilating pair. If the momentum space transform of
 $\psi(r)$ has a much smaller range of variation than that of the annihilation amplitude,
 the quasi--static approximation becomes applicable. In the
 opposite situation where these two scales of variation are similar, the
 quasi--static approximation can be seriously in error
 and needs to be replaced
 by the method of calculation outlined in Section 3 below.

 \subsection{2. The $K\bar K$ bound state}
    A non--relativistic approximation for the Bethe--Salpeter (BS)
  equation for weakly bound states \cite{LL74} was employed in \cite{KLS04} to
  study the mass and strong decays  of
  the $f_0(980)$ scalar meson considered
  as a bound kaon--antikaon pair of isospin zero interacting via vector meson exchange.
    In momentum space, this equation
  reads  
  \begin{eqnarray}
  \Big(\frac{{\bf p'\:^2}}{M_K}+2M_K-P_0\Big)\phi({\bf p'})=
   \frac{1}{4M_K^2}\int \frac{d^3p}{(2\pi)^3}
  \hat\Gamma({\bf p',p})\phi({\bf p})
  \label{e:BSm}                         
  \end{eqnarray}
  where $\hat\Gamma({\bf p',p})$ is the two--particle irreducible
  interaction kernel, or transition amplitude, appearing in  the BS equation that
  describes
  the non--relativistic scattering of the pair in the center of mass (CM) system 
  with a momentum change ${\bf  p} \to {\bf  p'}$ for either kaon of mass $M_K$.  

   The external four--momenta
  of the
  BS equation for bound states are off--shell but still obey
  four--momentum conservation. In particular the 
   sum of the initial, equal to final, time components appears explicitly as
   the energy eigenvalue parameter
   $P_0$ on the left hand side of 
   Eq.~(\ref{e:BSm}) and also  in the kernel
   $\hat\Gamma({\bf p,p'})=\hat\Gamma({\bf p,p'};P_0)$.
   This eigenvalue
  lies on the unphysical sheet of the complex energy plane where the four--momenta
  of the colliding particles are off their mass shell. It gives
   the mass and  decay half--width
   $P_0=M-i\Gamma/2 =2M_K+\varepsilon-i\Gamma/2$ of the bound system
   in the presence of
   interactions, where $\varepsilon < 0$ is the binding energy of the $K\bar K$
   molecule.
  In obtaining the non--relativistic form Eq.~(\ref{e:BSm}) it has to be
  assumed
  in addition  that these four--momenta become
   ``almost'' physical for weakly bound states \cite{LL74}.

   The corresponding eigenvalue
   equation in coordinate space is given by the Fourier transform of
   Eq.~(\ref{e:BSm}). This is just the usual Schr\"odinger wave equation
   containing a non--local, generally complex, potential with the
   bound state wavefunction $\psi({\bf r})$ having the transform
   $\phi({\bf p})$.
   In \cite{KLS04} an expression
   for $\hat\Gamma({\bf p',p})$ for
   $K\bar K$ scattering and
    annihilation has been derived
      using
     a standard $SU_V(3)\times
     SU_A(3)$ invariant interaction Lagrangian \cite{DESWART63} to describe vector meson
     exchange between the kaons in the non--relativistic limit. 
          In coordinate space this procedure leads to a real, 
     local one boson exchange potential, plus a pure imaginary contact term that describes the $K\bar K$
     annihilation into the $\pi\pi$ or $\pi\eta$  isoscalar and isovector channels
     respectively, i.e. to an optical potential of the form                                         
     \begin{eqnarray}
    {\cal V}_{opt}\psi({\bf r})=-\frac{1}{4M^2_K}
    \int d^3{\bf r'} \hat\Gamma({\bf r,r'})\psi({\bf r'})
    \approx-\Big[V_{OBE}^{(I)}(r)+\frac{i}{2}\lim_{v_r\to 0}
     (v_r\sigma^{(I)}_a)\;\delta^3({\bf r})\Big]\psi({\bf r}).  
     \label{e:vopt} 
     \end{eqnarray}
      Here $\hat\Gamma({\bf r,r'})$ is the coordinate space representation
      of $\hat\Gamma({\bf p',p})$ and $\sigma^{(I)}_a$ the  kaon--antikaon annihilation cross
      section for isospin $I$. 

     Explicit expressions for $V^{(0)}_{OBE}$ and $\sigma^{(0)}_a$ in the
     isoscalar channel that are relevant for the present discussion can be found
    in \cite{KLS04}. The resulting optical potential
   leads to a
   bound state solution for the $K\bar K$ system in an $s$--state
   that gives $P_0=(981-25i)$ MeV for the mass and  half
   width of $f_0(980)$ in good
   agreement with experiment \cite{PDG06}.
   While direct measurements of
   $K\bar K$ scattering and annihilation cross sections have not been
   reported against which the scattering predictions given by ${\cal V}_{opt}$ can be
   compared, one can do so indirectly by using
    two--channel unitarity to express the cross section for the inverse
    process $\pi^+\pi^-\to K\bar K$, 
    which has been measured \cite{SDP73}, in terms of 
    the common inelasticity
    parameter of the $K\bar K$ and $\pi\pi$ channels. This parameter may
    then be calculated
    from the $S$--matrix given by ${\cal V}_{opt}$ for the isoscalar channel, and 
    leads to a reasonable accord
    with  the experimental
    $\sigma(\pi^+\pi^-\to K\bar K)$  cross section \cite{KLS04}.

    Taken together, these facts suggest that the optical potential of
    Eq.~(\ref{e:vopt}) provides an adequate representation of the low energy dynamics
    of the $K\bar K$ system, at least for $I=0$.

 \subsection{3. Two--photon decay of $f_0(980)$  }

  We now revisit the two--photon decay width problem for $f_0(980)$.
  One can first implement the quasi--static approximation by replacing
  the  $v_r\sigma^{(0)}_a $ isoscalar cross section for annhilation into
  two pions with that for $K^+K^-$ annihilation into two photons
  and then calculating the resulting imaginary part
  $-i\Gamma(\gamma\gamma)/2$ of the energy shift given by Eq.~(\ref{e:vopt})
   in perturbation theory.
    The result is \cite{TB85,KLS04}

  \begin{eqnarray}
   \Gamma(\gamma\gamma)\approx \Gamma_{qs}(\gamma\gamma)=\frac{2\pi\alpha^2}{M^2_K}\psi^2_{K^+K^-}(0),
    \quad \psi_{K^+K^-}(0)=\frac{1}{\sqrt 2}\psi(0).
   \label{e:qstatic}
   \end{eqnarray}
 with $\alpha={e^2}/{4\pi}$, after projecting  out the
  charged channel amplitude given by the overlap
  $\zeta=<K^+K^-|K\bar K>=1/\sqrt 2$ for good isospin
   from the $K\bar K$ isoscalar ground state wave function $\psi(0)$ at the origin.
   In obtaining this estimate we have ignored a possible 
  $K^0\bar K^0\to 2\gamma$ contribution that lies beyond the scope of the
  present calculations that assume point kaons, but this contribution is known to be negligible 
  at low energies in any
  event \cite{TB85}.
     Using Eq.~(\ref{e:qstatic}) one finds  values for the annihilation
  width of $\sim 0.6$ to $0.9$ keV for the Barnes' potential depending
  on how $\psi(0)$ is estimated, or the much larger value
  $5.59$ keV for the one--boson exchange potential.
   The experimental value is listed as        
  $0.31\;^{+0.08}_{-0.11}$ keV \cite{PDG06}. 

    This spread of calculated values simply reflects the impact of the details of the assumed
  $K\bar K$ interaction potential on the value of $\psi(0)$ used in
  Eq.~(\ref{e:qstatic}). The more important question is
  how well Eq.~(\ref{e:qstatic}) actually meets the 
   quasi--static conditions used in its derivation, i.e. that of being able to calculate the annihilation while ignoring
   binding and vice versa, at the typical binding energies \cite{PDG06} of
   about $10$ to
   $20$ MeV
   encountered for the $K\bar K$ molecule.     
  In order to address this question we include the
  contribution of electromagnetic interactions
   between the charged kaons
    present
  in the $f_0(980)$ ground state \footnote{We ignore the
  instantaneous Coulomb attraction between the charged partners due to
  one--photon exchange, since this is insigificant in comparison with the
  strong interaction in the $K\bar K$ ground state. The roles of the strong
  and Coulomb interactions are
   reversed in the  $K^+K^-$ (kaonium) atom, where the Coulomb binding
   is dominant, the strong interaction endowing the pure Coulomb state
   with a small additional energy shift and  decay
   width \cite{KLS04}.}
     as part of
   the interaction kernel of the BS equation
   by replacing
    $\hat\Gamma({\bf p',p})$ with $\hat\Gamma({\bf p',p})+
    \hat\Gamma_{em}({\bf p',p})$ in Eq.~(\ref{e:BSm}). Since the
     $f_0(980)\to 2\gamma$ decay width is related to
    $Im\hat\Gamma_{em}$, we require two--particle diagrams with at least two
   intermediate state photons. For $K^+K^-$ scattering there are five
   diagrams of this type to order
   $\alpha^2$  that
   together form a gauge invariant set. They are shown in Fig.~1.

  Call $i\hat\Gamma_{em}$ the sum of these five diagrams.
  Then one can
  show after some calculation using the Cutkosky rules \cite{LL74}
  that the imaginary part of the electromagnetic contribution
  to Eq.~(\ref{e:BSm}) may be expressed in a form reminiscent of an
  optical theorem as
  \begin{eqnarray}
   2iIm\int \frac{d^3 p}{(2\pi)^3}\hat\Gamma_{em}({\bf p',p})\zeta\phi({\bf p})
     &=&\frac{i}{32\pi^2}
   \int \frac{1}{2}d\omega M^{\mu\sigma}({\bf p'})
   \int \frac{d^3 p}{(2\pi)^3}M_{\mu\sigma}({\bf p})\zeta\phi({\bf p})
  \nonumber
  \\
  &=&\frac{i}{32\pi^2}\int \frac{1}{2}d\omega
   \sum_{\lambda\lambda'}M^*_{\lambda\lambda'}({\bf p'})
  \int \frac{d^3 p}{(2\pi)^3}
   M_{\lambda\lambda'}({\bf p})\zeta\phi({\bf p})
   \label{e:tensor}
  \end{eqnarray}
  where $M_{\mu\sigma}({\bf p})$ is the tensor
    amplitude  for the annihilation of $K^+K^-$ into two photons
   at kaon momentum ${\bf p}$; $\zeta=1/\sqrt 2$ is the isospin projection factor as before.
  The symmetry factor $1/2$ arises naturally to restrict the integration
  $d\omega$ over the scattering direction of one of the photons
   to half the solid angle.
    The second form involving the sum over the photon 
    polarization vectors 
     $\epsilon^\mu_\lambda({\bf k})$, follows upon introducing their
     completeness relation and setting  
    \begin{eqnarray}
          M_{\lambda\lambda'}({\bf p})=
   [ \epsilon^\mu_\lambda({\bf k})M_{\mu\sigma}({\bf p})
    \epsilon^\sigma_{\lambda'}({-\bf k})]
   \label{e:matrix} 
   \end{eqnarray}

   We now obtain the electromagnetic decay width $f_0\to 2\gamma$ by including
   $Im\hat\Gamma_{em}({\bf p',p})$ in the right hand side of
   Eq.~(\ref{e:BSm}) for the charged channel and calculating
   the additional imaginary shift in the total energy
   $P_0\to P_0-i\Gamma(\gamma\gamma)/2$ that this produces in perturbation theory.
 Then
 \begin{eqnarray}
 \Gamma(\gamma\gamma)
 &=&\frac{1}{4M^2_K}\int\int\frac{d^3p'}
 {(2\pi)^3}
 \frac{d^3p}{(2\pi)^3}\zeta\phi^*({\bf p'})[2Im\hat\Gamma_{em}({\bf p',p})]
 \zeta\phi({\bf p})
  \nonumber
  \\
 &=&\frac{1}{64\pi M^2_K}\sum_{\lambda\lambda'}|\int
  \frac{d^3p}{(2\pi)^3}\zeta\phi({\bf p})M_{\lambda\lambda'}({\bf p})|^2;
 \label{e:gammafnl}
  \end{eqnarray}
  Note in passing that this form reduces to the quasi--static result
 in Eq.~(\ref{e:qstatic}) if only the static part ${\bf p}\to 0$
  of the annihilation
 amplitude is kept. For then $M_{\lambda\lambda'}$ factors out of the momentum
 integral to reproduce the free annihilation cross section as
 $v_r\sigma_{em} =\sum_{\lambda\lambda'}|M_{\lambda\lambda'}|^2/(64\pi M^2_K)$,
while the remaining  integral squared over $\zeta\phi({\bf p})$ just equals $\psi^2_{K^+K^-}(0)$. 

Since the Fourier transforms $\phi({\bf p})=\phi(p)$ are spherically symmetric
in  momentum space for $s$--waves it is convenient to split the integral in
Eq.~(\ref{e:gammafnl}) 
into radial and angular parts,                         
\begin{eqnarray}
 &&I_{\lambda\lambda'}=\int
 \frac{d^3p}{(2\pi)^3}\phi({\bf p})M_{\lambda\lambda'}({\bf p})=
 \frac{1}{(2\pi)^3}\int_0^\infty dp p^2\phi(p)\Theta_{\lambda\lambda'}(p)
\end{eqnarray}
where
\begin{eqnarray}
&&\Theta_{\lambda\lambda'}(p)= \int d\Omega
 M_{\lambda\lambda'}({p,\Omega})
\label{e:ang} 
\end{eqnarray}

 The transition amplitudes $M_{\lambda\lambda'}({\bf p})$ in these expressions are required at
  off--shell values of the
  kaon four--momenta. We take these  as
   $[\frac{1}{2}P_0,{\bf \pm p}]\approx [M_K,{\bf \pm p}]$ in the CM system after ignoring the
 small binding energy of the kaon pair. The outgoing photons
 are of course still on--shell with four--momenta  $[M_K,{\bf \pm k}]$ 
  where $|{\bf k}|=M_K$ to the same approximation; each photon carries away
 half the total mass of the  decaying bound state.
We evaluate $M_{\lambda\lambda'}({\bf p})$ at these off--shell four--momentum values
from the standard expression for the annihilation amplitude 
$M_{\mu\sigma}({\bf p})$ 
for charged point-like bosons, without form factors, that can be found
in \cite{IZ86} for example. Then from Eq.~(\ref{e:matrix})
%as the sum of
%``seagull'' plus direct and crossed diagrams. One finds
 
 \begin{eqnarray}
M_{\lambda\lambda'}({\bf p})=
e^2\Big\{2{(\bf e_\lambda\cdot\bf e_{\lambda'})}
+\frac{(2{\bf p}\cdot {\bf e}_\lambda)
(-2{\bf p}\cdot {\bf e}_{\lambda'})}{M_K^2+({\bf p-k})^2}
+ ({\bf k\to -k})\Big\} 
 \end{eqnarray}
where  $|{\bf k}|=M_K$,  after introducing the transverse gauge $\epsilon^\mu_\lambda({\bf k})=
[0,{\bf e}_\lambda]$ with ${\bf k}\cdot {\bf e}_\lambda=0$
and ${\bf e}_\lambda\cdot {\bf e}_{\lambda'}=\delta_{\lambda\lambda'}$
for the photon polarization vectors.

 The angular
integral in Eq.~(\ref{e:ang}) over all kaon momentum directions 
 is  easily
performed relative to a set of axes where the direction of the photon
momentum ${\bf k}$
defines
 the polar axis and 
 the two remaining orthogonal directions by
 the two polarization vectors of either photon. Then
$\Theta_{\lambda\lambda'}(p)$ has a common diagonal value in the polarization
indices  and is given by
 \footnote{For comparison, using on--shell 
 instead  of the off--shell kaon four--momenta replaces
 $\theta_{\gamma\gamma}(p)$ by $[p(2+p^2)]^{-1}\ln [(2+p^2+2p)/(2+p^2-2p)]$ that only
 starts to differ from  Eq.~(\ref{e:theta}) at ${\cal O}(p^4)$.}

 \begin{eqnarray}
 \Theta_{\lambda\lambda'}(p)= 8\pi e^2\theta_{\gamma\gamma}(p)\delta_{\lambda\lambda'},\quad
\theta_{\gamma\gamma}(p)= \Big[\frac{4+p^4}{8p}\ln\frac{2+p^2+2p}{2+p^2-2p}-\frac{1}{2}p^2\Big]
 \label{e:theta}
 \end{eqnarray}      
  after expressing the  three--momentum $p$ 
  in units of $M_K$.
    Thus the final integral $I_{\lambda\lambda'}$ that is required for the width
 calculation is also diagonal, with a common value $I$ that we factor as
 \begin{eqnarray}
 I= 8\pi\alpha\psi(0)R,\quad 
R=\int_0^\infty dp f(p)\theta_{\gamma\gamma}(p),
\label{e:R}
\end{eqnarray}
 with $f(p)$ defined as the normalized Fourier transform 
 \begin{eqnarray}
 f(p)= \frac{p^2}{2\pi^2\psi(0)}\phi(p);\quad \int_0^\infty dpf(p)=1.
 \label{e:f}
 \end{eqnarray}          
As before $\psi(0)$  is
the radial wave function of the isoscalar $K\bar K$ pair at the origin. Thus
 $R=1$ if
$\theta_{\gamma\gamma}(p)$
 is replaced by its static limit value $\theta_{\gamma\gamma}(0)=1$.
 The final form of Eq.~(\ref{e:gammafnl})
 for the two--photon decay width then reads
\begin{eqnarray}
&&\Gamma(\gamma\gamma)=\frac{2\pi\alpha^2}{M_K^2}\psi^2_{K^+K^-}(0)R^2=
 \Gamma_{qs}(\gamma\gamma)R^2
\label{e:gammaR}
\end{eqnarray}
The coefficient  $R^2$ directly measures the deviation of this width from
the quasi--static
limit given by Eq.~(\ref{e:qstatic}). 

\subsection{4. Calculations for two different potentials}

 Let us now recalculate the two--photon decay width 
 for both the one--boson exchange (OBE) potential and
Barnes' gaussian potential.
For computational convenience it is useful
 to replace both potentials by their Bargmann equivalent potentials \cite{VB49} that have the
 same  scattering length $a_0$ and effective range $r_0$ as calculated
 numerically for the original potential forms. Then ($a_0,r_0$)
  translate into the two parameters $(a,b)$  of the equivalent Bargmann potential form
 given in the Appendix.
 Analytic expressions for both the bound state
  wave function  as well as its Fourier transform are then available in closed
  form, see Eqs.~(\ref{e:bwf}), (\ref{e:fourier}) and  Fig.~\ref{f:fig3}.

(i) {\it One Boson Exchange Potential}. For this case the
scattering length and effective range 
 were calculated in \cite{KLS04}. They are $a_0=5.835M_K^{-1}$
 and $r_0=1.187M_K^{-1}$, giving
  $a=-0.1936M_K$ and $b= 1.491M_K$.
   For this parameter set there is a single
 bound state at $-18.60$ MeV. Then  Eq.~(\ref{e:bwf}) gives
 $\psi(0)=0.260M^{3/2}_K$, that leads to $\Gamma_{qs}(\gamma\gamma)=5.59$ keV. Taking $\phi(p)$ from Eq.~(\ref{e:fourier})
 to do the integral for $R$ numerically, one finds $R=0.287$ so that
 \begin{eqnarray}
 \Gamma(\gamma\gamma)=0.46^{+0.10}_{-0.13}\;{\rm keV},
 \end{eqnarray}                                                
 or a reduction in the  quasi--static value
  by  more than an order of magnitude.  The error bars have been estimated
  by considering the
   experimental uncertainties \cite{PDG06} on
   $P_0=[(980\pm 10) -i(20\; {\rm to}\; 50)]$ MeV for the $f_0(980)$
   mass and  $2\pi$ decay half width that change the values of $a$ and
  $b$ accordingly \cite{KLS04}.

  (ii) {\it Gaussian Potential}. In \cite{TB85} Barnes  estimated $\psi(0)$ using a gaussian potential form
  $V(r)=-V_0exp(-r^2/2r^2_{g})$ with $V_0=440$ MeV fitted to the typical
  interkaon potential depth of the Weinstein--Isgur model, and a 
   range of $r_g=0.57$ fm $=1.435 M^{-1}_K$ in order to reproduce
   a binding
   energy of $\sim -10$ MeV for the $K\bar K$ pair. The  scattering
   length and effective range for this potential are 
   nearly double those of the OBE potential: one calculates that
   $a_0=8.404M_K^{-1}$, $r_0=2.397M_K^{-1}$  giving $a=-0.1437M_K$,
   $b=0.6910M_K$ for the parameters of
   its Bargmann equivalent. 
   The  larger scattering length translates into a weaker
   binding energy, $-10.2$ MeV,
    and thus a smaller wave function at the origin,
   $\psi_g(0)=0.102M^{3/2}_K \alt \frac{1}{\sqrt{6}}\psi(0)$.
   This gives the much smaller
   value $ 0.86$ keV for $\Gamma_{qs}(\gamma\gamma)$. 
   Doing the $R$ integral as under (i) the result is
   $R=0.632$ so that now
 \begin{eqnarray}
 \Gamma(\gamma\gamma)=0.34\;{\rm keV}
 \end{eqnarray}                                                

  This is only about $2 \frac{1}{2}$ times smaller than $\Gamma_{qs}(\gamma\gamma)$.
  The quasi--static
  limit is thus more reliable for kaons moving in the Barnes' gaussian
  potential than in the
  one boson exchange potential. The reason for this is clear from
  Fig.~\ref{f:fig2}. This figure compares the normalized Fourier transforms
   $f(p)$ of Eq.~(\ref{e:f}) for both potentials
  over the  momentum range $0<p<3$ GeV of relevance for the present
  calculation. The gaussian has
  a smaller depth and longer range than the OBE potential. Thus the  $f(p)$
   that determines $R$ embraces a much
    smaller range of kaon momenta
    for the Barnes' potential than it does for the OBE potential, and so gives
   an $R$ less different from unity in the first case.
  Moreover, by equating the binding energy of the Barnes potential to that of the
  OBE potential without changing its range one finds that it predicts a $\Gamma(\gamma\gamma)=0.44$ keV, very
  close to the OBE result. Similarly, the OBE potential with its binding set equal to
  that of the Barnes' potential leads to $\Gamma(\gamma\gamma)=0.38$ keV
  that in turn
  lies close to the Barnes' result. Thus the calculated two-photon width depends
  mainly on the binding energy of the decaying state. 
  This is a general feature of the weak binding limit $\kappa^2<<b^2$.
   For then it follows from Eqs.~(\ref{e:bwf}), (\ref{e:fourier}) and (\ref{e:f})
    that
    $\Gamma_{qs}\sim \psi^2(0)\sim {\kappa}b^2$
     depends on  both the
    binding energy and potential range parameters $\kappa$ and $(2b)^{-1}$,
    while
    the factor $R$ of
   Eq.~(\ref{e:gammaR}) varies mainly with the range like
     $R\sim 1/b$. This is so because its integrand behaves like $1/b^2$
   times a function of $p$ that suppresses
   the momentum integration beyond $p\sim 2b$.
   Consequently the range parameter  cancels out upon forming the product
   $\psi^2(0)R^2$,
   leading to the behavior
   $\Gamma(\gamma\gamma)\sim \kappa \sim \surd (-\epsilon)$.

      The results are summarized in  Fig.~\ref{f:fig2} and Table I respectively.
      The conclusion is thus that there is little to distinguish between these two
   potential models for describing the $K\bar K$ molecule as far  the
   two--photon decay is concerned, once the momentum distribution of
   relative motion of the interacting
    pair is taken into account. Both lead to  very similar two--photon decay
    widths
    that are in fact in semi--quantitative agreement with experiment.

  \subsection{5. Comparison with $\pi\pi$ decay of $f_0(980)$}
    
   The calculated value of $\Gamma(\pi\pi)\approx 50$ MeV for the $f_0\to 2\pi$
   decay width already quoted in Section 2
   for the OBE
   potential model  using the quasi--static approximation
  falls well within the rather wide range
  of experimental values for $\Gamma(\pi\pi)$ of $40$ to $100$ MeV
    reported in the literature \cite{PDG06}. On the other hand the
    gaussian model scales down
    this result  
    by $\psi^2_g(0)/\psi^2(0) \sim 1/6$ 
   to $\sim 8.3$ MeV that falls way below experiment.
        It is therefore important to check
    whether non--static effects can also lead to significant changes in either of these
    estimates. 

    We assume as before \cite{KLS04} 
   that 
   the transition amplitude for the two-pion decay of the $f_0(980)$
    proceeds via  
   $K^*$ vector meson $t$--channel exchange \footnote{
   Additional contributions to the $K\bar K\to \pi\pi$ width can
    arise from $s$--channel scalar
    meson exchange.
   Using the relevant coupling constants
   derived in \cite{RD98} from the linear $\sigma$ model for the exchange of the
   $m_\sigma \approx 600$ MeV $f_0(600)$, or
   $\sigma$ meson, 
    one estimates a 
   contribution of $\alt 11$ MeV to the total decay width from this source.
   This small value is mainly due to  suppression of the $K\bar K$ to 
   nonstrange $\sigma$ coupling.}.
   Then the photons are replaced by pions and
   the intermediate kaons by $K^*$'s in the first two diagrams of Fig.~1.
   Note that both
   the direct and crossed diagrams  contribute since the
   the pions are identical bosons in the isospin basis.  The imaginary part of this
   combination leads to the analog of Eq.~(\ref{e:tensor}) for pions with the
   tensor scattering amplitude $M_{\mu\sigma}$ replaced by a
   scalar amplitude $M_{\pi\pi}$. 
     Then
    Eq.~(\ref{e:theta}) for pions reads
   $\Theta_{\pi\pi}(p)=8\pi g^2_{\pi\pi K\bar K}\theta_{\pi\pi}(p)$ where
   $g^2_{\pi\pi K\bar K}$  is an effective coupling constant that determines
   the free $K\bar K\to \pi\pi$ annihilation amplitude. Its value is not needed for the present discussion.
    Omitting the calculational details, one finds 
    \begin{eqnarray}
    \theta_{\pi\pi}(p)=\Big(\frac{M_{K^*}^2+p^2_\pi}{4+p^2_\pi}\Big)
     \Big[\frac{4+M_{K^*}^2+2p^2_\pi+2p^2}{4pp_\pi}
     \ln\frac{M_{K^*}^2+(p+p_\pi)^2}{M_{K^*}^2+(p-p_\pi)^2}-1\Big]
      \label{e:thetapi}
      \end{eqnarray}
     in units $M_K$, that simplifies to
       \begin{eqnarray}
       \theta_{\pi\pi}(p)\approx \Big[\frac{5+p^2}{2p}
       \ln\frac{5+p^2+2p}{5+p^2-2p}-1\Big]
      \end{eqnarray}
       if we use the estimates
      $M_{K^*}\approx 2M_K$ and $p_\pi\approx M_K$ for the $K^*$ mass and
       pion threshold momentum in order to compare more directly with
      $\theta_{\gamma\gamma}(p)$.
      The function $\theta_{\pi\pi}(p)$ is shown in Fig.~2.
      Replacing $\theta_{\gamma\gamma}(p)$ by either form for
       $\theta_{\pi\pi}(p)$ in
       Eq.~(\ref{e:R}), one obtains a  modification factor of
       $R_{\pi\pi} \alt 1.1$  for both the OBE and gaussian potentials, an
       insignificant change. One can understand this result without any
       calculation. It comes about because the
       momentum scale over which  $\theta_{\pi\pi}(p)$ varies
       is fixed by the mass of the exchanged boson, in this case the $K^*$,
       that is nearly a factor two more massive than the exchanged kaon that
       fixes the momentum scale for $\theta_{\gamma\gamma}(p)$. This means that the
       value of the integral determining  $R_{\pi\pi}$  becomes almost the
       same as the
       normalization integral for $f(p)$ since $\theta_{\pi\pi}(p)$ hardly varies
       at all over the momentum range of $f(p)$. In contrast with the $2\gamma$ decay
       problem, the quasi--static approximation
       introduces an insignificant error into the calculation of the $2\pi$ decay
       width. The entries for $\Gamma(\pi\pi)$ in Table I are thus essentially unchanged from their
       quasi--static values.

       \subsection{6. Discussion and conclusions}

        We have shown that calculating the annihilation width of a decaying
        bound system in the quasi--static approximation can be seriously
        in error when the momentum ranges of
        variation of the annihilation amplitude and the momentum transform of
        the bound state wave function are similar, and have given a revised
        formulation by including electromagnetic corrections in the kernel of the
         Bethe--Salpeter equation to cover this case. The resulting changes in  the calculated two--photon
        annihilation widths  for  $f_0(980)\to \gamma\gamma$ in the
        molecular picture
          are considerable for one boson exchange model,
         less so for the Barnes' potential model. In fact the predictions
         for these two competing  potentials are brought
         into semi--quantitative agreement with each other and experiment
         as shown in Table 1.
         In contrast the calculated two--pion annihilation width for $f_0(980)\to
         \pi\pi$
          is essentially  unaffected by non--static corrections,
          but has
         completely different values for the two models that  favor
         the one boson exchange model.

         Since the two--photon decay  width
         in the OBE model has been decreased by nearly an order
         of magnitude by the non--static corrections
          to already agree semi--quantitatively  with experiment, the
         speculation in Ref.~\cite{KLS04} of important
         admixtures of, for example, $q^2\bar q^2$ states
         in the pure $K\bar K$ ground state
         in order to reproduce the two--photon width based on the
         quasi--static approximation  falls away: The
          one boson exchange model  reproduces the mass, as well as reasonable
          two--pion and
         two--photon decay widths for  $f_0(980)$ 
         without any
         further assumptions once non--static effects are included in
         the electromagnetic sector, thereby giving additional support to
         the suggestion that this meson is predominantly a $K\bar K$ molecule.

\subsection{7. Acknowledgments}

     This reseach was  supported
    by the Ernest Oppenheimer Memorial Trust
    in terms of  a Harry Oppenheimer Fellowship Award, which is gratefully
    acknowledged. 
    I would also like to thank Professor Helmut Hofmann and  other members of the
    Theory Division at the Physik--Department of the Technische Universit\"at
    M\"unchen at Garching for their kind hospitality, as well as
    Dr Veljko Dmitra$\check{s}$inovi$\acute{c}$ of the Vin$\check{c}$a
    Institute of Nuclear Sciences, Belgrade, for essential correspondence.

\subsection{8. Appendix: Equivalent Bargmann Potentials}
  For calculational simplicity we replace both the expression for 
  $V_{OBE}$ that appears in Eq.~(\ref{e:vopt}) as well  as Barnes' gaussian potential
  by their Bargmann
 equivalents that have the same respective scattering lengths and effective
 ranges. These are given by the two--parameter potential \cite{VB49}
    \begin{eqnarray}
   V_{Barg}(r)=\frac{1}{M_K}\frac{8b^2}{b^2-a^2}\Big[\frac{e^{br}}{b-a}
    +\frac{e^{-br}}{b+a}\Big]^{-2}
    \label{e:Barg}
    \end{eqnarray}
    with $a=-[1-\sqrt{1-2r_0/a_0}]/r_0$ and $b= [1+\sqrt{1-2r_0/a_0}]/r_0$
    in order to reproduce the same scattering length
    $a_0$ and effective range $r_0$ of the original potential.

    If $a=-\kappa<0$ the
    Bargmann potential in Eq.~(\ref{e:Barg}) has a single
    bound $s$--state at  energy
    $\epsilon=-\kappa^2/M_K$.
    The  normalized bound state
    eigenfunction belonging to this eigenvalue is given by 
    \begin{eqnarray}
     &&\psi(r)=\psi(0)\frac{u(r)}{r},\quad
     \psi(0)=\Big[\frac{2\kappa(b^2-a^2)}{4\pi}\Big]^{1/2}
     \nonumber
     \\
     &&u(r)= \frac{\tanh(br)}{b-\kappa\tanh(br)}e^{-\kappa r}\to r,\quad r\to 0
     \label{e:bwf} 
     \end{eqnarray}
      We note that for weak binding in a short range potential,
      i.e. $a_0>>r_0$, the Bargmann potential
      parameters below Eq.~(\ref{e:Barg}) reduce  to $a\approx -1/a_0$ and
      $b\approx 2/r_0$ approximately. In this limit
      one retrieves the standard universal forms \cite{EBMK04}
      $\epsilon\approx -1/M_K a_0^2$
     and $r\psi(r)\approx (2\pi a_0)^{-1/2}exp(-r/a_0)$
     for the binding energy and
      radial wavefunction
      at large distances $r>>r_0$.

      We also require the Fourier transform $\phi({\bf p})$ of $\psi({\bf r})$ defined by
       \begin{eqnarray}
       \psi({\bf r})=\int \frac{d^3p}{(2\pi)^3}\phi({\bf p}) e^{i\bf p\cdot r}
       \end{eqnarray}
       This only
      depends on the magnitude of the momentum $p=|{\bf p}|$ since
       $\psi({\bf r})$ in Eq.~(\ref{e:bwf}) is spherically symmetric. One finds
                                           
      \begin{eqnarray}
      \phi(p)&=&\int d^3{\bf r}\psi(r)e^{-i{\bf p\cdot r}}=
      \frac{4\pi}{p}\psi(0)Im\int_0^\infty dr\;u(r)e^{ipr}
      \nonumber
      \\
     &=&\frac{4\pi}{p}\psi(0)\frac{1}{4b^2}
     Im \Big\{\frac{1}{\rho(1+\rho)}\;
     {_2} F_{1}[1,2,2+\rho;\frac{1}{2}+\frac{\kappa}{2b}]\Big\},
      \quad \rho=\frac{\kappa}{2b}-i\frac{p}{2b}
      \nonumber
      \\
    &\sim& -\frac{1}{p^6},\quad p\to \infty
      \label{e:fourier}
      \end{eqnarray}
      in terms of the hypergeometric function
      ${_2} F_{1}(\alpha,\beta,\gamma;z)$ \cite{AS}.
            This result follows immediately upon
      using the substitution $x=1-exp\;(-2br)$ to transform the radial integral into
      an integral representation of the hypergeometric function.

\newpage
 
\begin{table}[ht]
\caption{\label{t:table1}
Summary of the calculated values for the one--boson exchange and gaussian
potentials. The numbers quoted below have been obtained
from their Bargmann equivalent potentials as described in the text.
The experimental
values from various sources have also been listed.}

\begin{ruledtabular}
\begin{tabular}{lccccc}
Source   & 
$\psi(0)$ GeV$^{3/2}$
& $ \Gamma_{qs}(\gamma\gamma)$ keV 
& $R_{\gamma\gamma}$ & $\Gamma(\gamma\gamma)$ keV & $\Gamma(\pi\pi)$ MeV\\
\colrule
\\
 One boson exchange potential \cite{KLS04} &  $0.091$ & $5.59$
 & $0.287$
 & $0.46^{+0.10}_{-0.13}$ & $\sim 50$\\
 \\
 Gaussian potential \cite{TB85} & $0.036$ & $ 0.86$
  & $0.632$
 &  $0.34$ & $\sim 8.3$\\ 
 \\
 Particle Data Group \cite{PDG06} &--&--&--& $0.31^{+0.08}_{-0.11}$
 & $ 40$ to $100$\\
 \\
  Fermilab E 791 Collaboration \cite{E791} &-- &-- &--&--
  & $44\pm 4$ \\

 \end{tabular}
\end{ruledtabular}
\end{table}
\newpage

 \begin{figure}
%\resizebox{0.9\textwidth}{!}{\rotatebox{-90}{\includegraphics
%{bind0.eps}}}%
\rotatebox{0}{\includegraphics{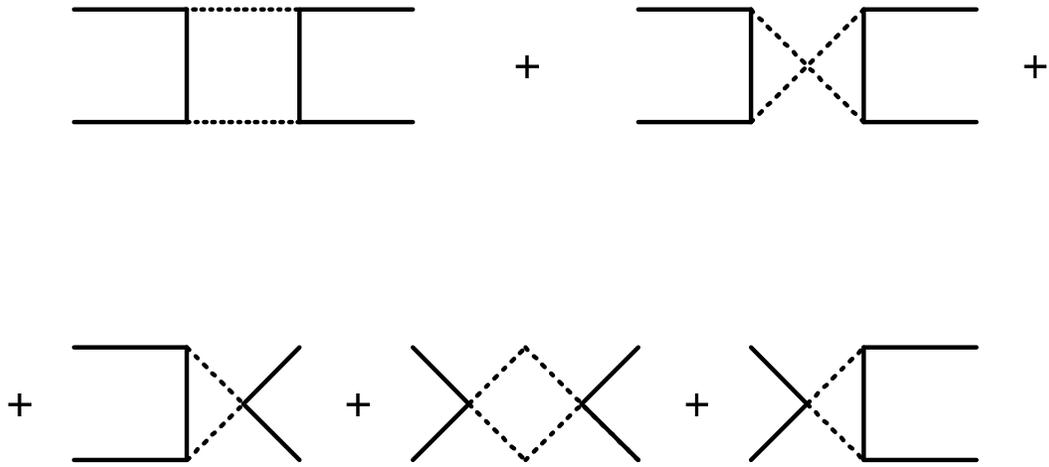}}
\caption{\label{f:fig1}
  The five diagrams that contribute  to order $\alpha^2$ to the
  electromagnetic part of interaction kernel in the Bethe--Salpeter equation 
    for
  $K^+K^-$ scattering. Solid lines: charged kaons, broken lines: photons.}

\end{figure}

 \begin{figure}
%\resizebox{0.9\textwidth}{!}{\rotatebox{-90}{\includegraphics
%{bind0.eps}}}%
\rotatebox{0}{\includegraphics{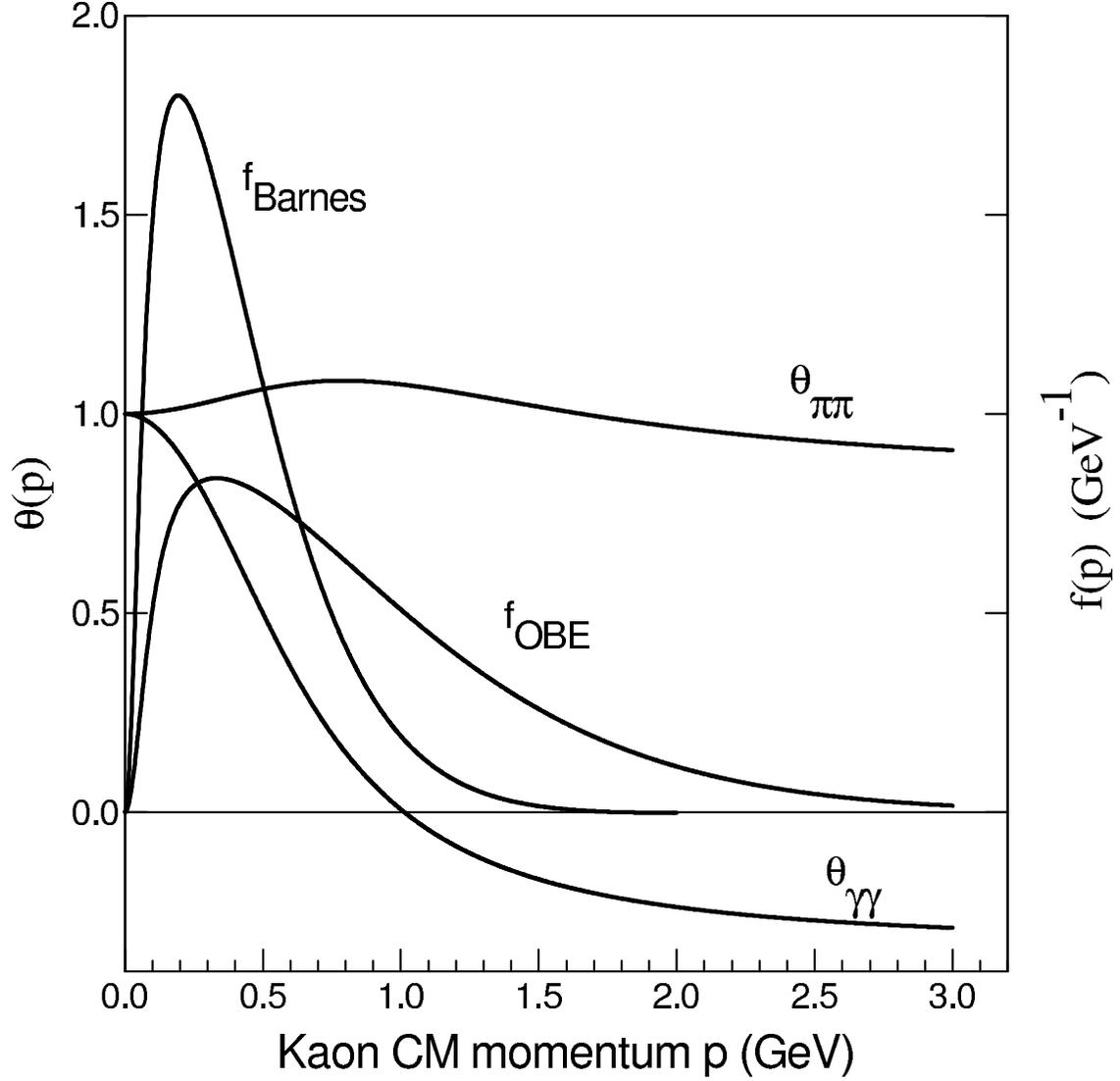}}
\caption{\label{f:fig2}
  Variation of the angular integrals $\theta(p)$ given by Eqs.~(\ref{e:theta})
  and (\ref{e:thetapi}) with the kaon CM momentum for $2\gamma$ versus $2\pi$ decay of
  the $f_0(980)$ (left hand scale). Also shown are the corresponding
  normalized Fourier transforms
  $f(p)=p^2\phi(p)/(2\pi^2 \psi(0))$ of Eq.~(\ref{e:f}) for the one boson exchange (OBE)
  and Barnes' gaussian potentials respectively
  (right hand scale).}

\end{figure}

 \newpage

\begin{figure}
%\resizebox{0.9\textwidth}{!}{\rotatebox{-90}{\includegraphics
%{bind0.eps}}}%
\rotatebox{0}{\includegraphics{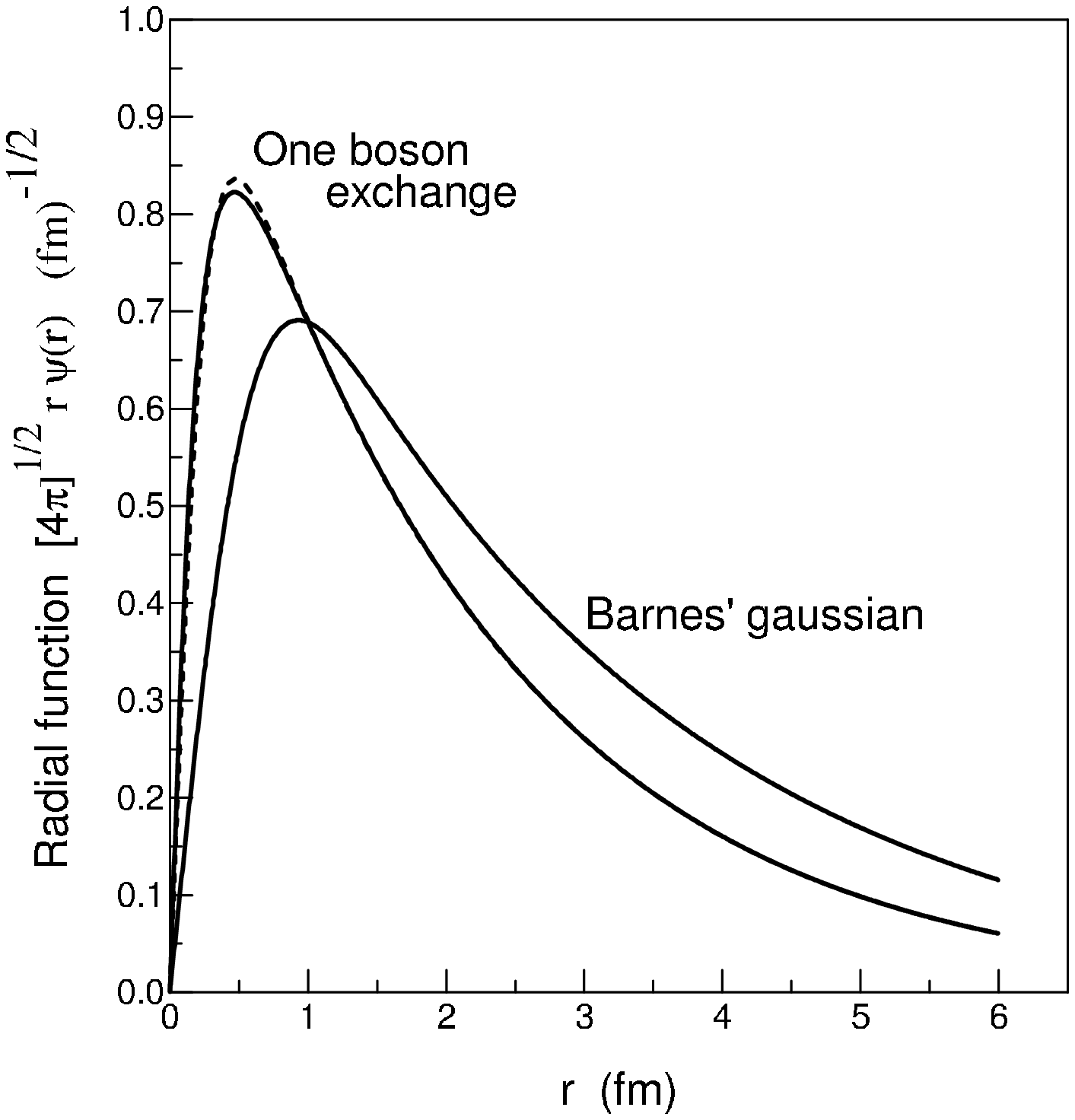}}
\caption{\label{f:fig3}
 Comparison of the numerically generated  normalized radial bound state wave functions for the
 one--boson exchange and
 Barnes' gaussian potentials (solid curves) respectively, with their analytic counterparts given by
 Eq.~(\ref{e:bwf})
 for their equivalent Bargmann potentials (broken curves). The numerical and
 analytic wave functions
  essentially coincide in the case of the Barnes' potential.}
 \end{figure}


\begin{thebibliography}{17}




\bibitem{LM00}

L. Montanet, Nucl. Phys. B (Proc. Suppl.) {\bf 86} (2000) 381; V.V. Anisovich
{\it et al.}, Phys. Lett. B {\bf 480} (2000) 19.

\bibitem{NNA89}

N. N. Achasov and V. N. Ivanchenko, Nucl. Phys. B {\bf 315} (1989) 465.


\bibitem{WIsgur82}
J. Weinstein and N. Isgur, Phys. Rev. Lett. {\bf 48} (1982) 659 ; Phys. Rev.
{\bf D 27}, 588 (1983); {\bf D 41} (1990) 2236.

\bibitem{KLS04}                         
S. Krewald, R. H. Lemmer and F. P. Sassen, Phys. Rev. D {\bf 69} (2004) 016003.

\bibitem{RKNI87}

R. Kokoski and N. Isgur, Phys. Rev. D {\bf 35} (1987) 907.

\bibitem{RT02}

R. H. Lemmer and R. Tegen, Phys. Rev. C {\bf 66} (2002) 065202.

 
\bibitem{PDG06}
W.--M. Yao {\it et al.}, (Particle Data Group),
J. Phys. {\bf G 33} (2006) 1.


\bibitem{TB85}

T. Barnes, Phys. Lett. {\bf 165 B} (1985) 434.

\bibitem{MAS75}

J. A. Wheeler, Ann. NY Acad. Sci. {\bf 48} (1946) 219;
M. A. Stroscio, Phys. Rep. {\bf C 22} (1975) 215.



\bibitem{LL74}

See, for example, L. D. Landau and E. M. Lifshitz, {\it Relativistic quantum
theory}, Course of Theoretical Physics Vol. 4 (Pergammon, Oxford, 1974).


\bibitem{DESWART63}
J. J. de Swart, Rev. Mod. Phys. {\bf 35} (1963) 916.






\bibitem{SDP73}
S. D. Protopopescu {\it et al.}, Phys. Rev. {\bf D 7} (1973) 1279 ; A. D. Martin
{\it et al.}, Nucl. Phys. {\bf B 121} (1977) 514 ; B. Hyams {\it et al.}, Nucl.
Phys. {\bf B 64} (1973) 134.


\bibitem{IZ86}

C. Itzykson and J. B. Zuber, {\it Quantum Field Theory}, (McGraw--Hill, New York,
1986).

\bibitem{VB49}
V. Bargmann, Rev. Mod. Phys. {\bf 21} (1949) 488.


\bibitem{RD98}
R. Delbourgo and M.D. Scadron, Int. J. Mod. Phys. A{\bf 13} (1998) 657;
J. Schechter and Y. Ueda, Phys. Rev. D{\bf 3} (1971) 2874.





\bibitem{EBMK04}
 E. Braaten and M. Kusunoki, Phys. Rev. {\bf D 69} (2004) 074005;
 M. B. Voloshin, Phys. Lett. {\bf B 579} (2004) 316.



\bibitem{AS}
 See, for example, M. Abramowitz and I. A. Stegun, Editors, {\it Handbook of Mathematical
 Functions} (Dover Publications, Inc., New York, 1965).



\bibitem{E791}

Fermilab E791 Collaboration, E. M. Aitala {\it et al.}, Phys. Rev. Lett.
{\bf 86} (2001) 765.
 \end{thebibliography}
\end{document}